\documentclass[final,5p,times,twocolumn]{elsarticle}
\usepackage{graphicx}
\usepackage{slashed,color}
\usepackage{amsmath,amsfonts,bm}
\usepackage{amssymb}
\usepackage[T1]{fontenc}

\newcommand{\nn}{\nonumber \\ }

\journal{Physics Letters  B}

\begin{document}



\author{A.~V.~Radyushkin}
\address{Physics Department, Old Dominion University, Norfolk,
             VA 23529, USA}
\address{Thomas Jefferson National Accelerator Facility,
              Newport News, VA 23606, USA
}

\title{Structure  of parton    quasi-distributions and their moments}

\begin{abstract}

We discuss the structure of the  parton    quasi-distributions (quasi-PDFs) $Q(y, P_3)$ outside 
the  ``canonical''   $-1 \leq y \leq 1$  support region of  the  usual parton distribution functions (PDFs). 
Writing the $y^n$ moments of $Q(y, P_3)$ in terms of the combined \mbox{$x^{n-2l} k_\perp^{2l}$-moments} 
of the transverse momentum distribution (TMD) ${\cal F} (x,k_\perp^2)$,
we establish a connection between the large-$|y|$   behavior of $Q(y,P_3)$ and
large-$k_\perp^2$ behavior of ${\cal F} (x,k_\perp^2)$. 
In particular, we show that the $1/k_\perp^2$ hard tail of TMDs in QCD results 
in a slowly decreasing $\sim 1/|y|$  behavior of quasi-PDFs for large $|y|$ that produces infinite $y^n$
moments of $Q(y,P_3)$.  We also relate the $\sim 1/|y|$ terms with the $\ln z_3^2$-singulariies 
of the Ioffe-time pseudo-distributions $\mathfrak{M} (\nu, z_3^2)$.
   Converting the operator product expansion for $\mathfrak{M} (\nu, z_3^2)$
   into a matching relation between the quasi-PDF $Q(y,P_3)$ and the 
   light-cone PDF $f(x, \mu^2)$, we demonstrate that there is no contradiction 
   between the infinite values of the $y^n$ moments of $Q(y,P_3)$ and 
   finite values of the $x^n$ moments of $f(x, \mu^2)$.

\end{abstract}

\maketitle


\section{Introduction}

 In the original Feynman approach \cite{Feynman:1973xc},  the parton distribution functions (PDFs)  $f(x)$
  were introduced as the   infinite momentum $P_3 \to \infty$ limit 
  of  distributions  in the longitudinal 
$k_3 = y P_3$ momentum of partons. 
These distributions   basically coincide 
with  the  quasi-PDFs   $Q(y, P_3)$ introduced more recently by X. Ji \cite{Ji:2013dva}. 

As is well-known, ``$x$'' of the parton model 
corresponds to the ratio $x=k_+/P_+$ of the light-cone-plus  
components of the parton and hadron momenta,
rather than the ratio $y=k_3/P_3$ of their third Cartesian components.
However, in the $P_3 \to \infty$ limit, the difference 
between $y$ and $x$ disappears.

In the parton model, 
 $f(x)$'s   were treated as $k_\perp$-integrals of more detailed  $f(x, k_\perp)$ distributions 
 that involve  also the transverse momentum $k_\perp$. 
 From the start,   it was understood  by Feynman  that   the 
 $P_3 \to \infty$
  limit exists   only if $f(x, k_\perp)$ 
 rapidly decreases  with  $k_\perp$, so that the integral over $k_\perp$
 does not diverge. 
This happens, in particular,    in the theories/models  with transverse 
momentum cut-off $k_\perp \lesssim \Lambda$, e.g.,  in super-renormalizable  models,
but not in QED and other renormalizable field theories.

 One may ask   two natural questions. First,  
why 
the shape of $Q(y,P_3)$
 for a finite $P_3$  differs from 
 that of
 $f(x)$?
 Second, how  does 
the shape of $Q(y,P_3)$
convert  into  
 that of
 $f(x)$ when $P_3 \to \infty$? 
 A qualitative answer  is that the parton's  longitudinal momentum $k_3=y P_3$ comes from two sources:  
 from  the motion of the hadron  as a whole  ($xP_3$) and  
from a  Fermi motion of quarks inside the hadron, so that   $(y-x)P_3 \sim 1/R_{\rm hadr}$.
As \mbox{$P_3 \to \infty$,}  the role of the $y-x \sim 1/P_3R_{\rm hadr}$ fraction 
decreases and  
$Q(y,P_3) \to f(x)$.

In this picture,  the 
 $(y-x)P_3$  part has the same  physical 
 origin as the parton's  transverse momentum. 
Hence, one should be able to relate quasi-PDFs  to the transverse
momentum distributions   (TMDs)  and quantify the difference between $Q(y,P_3)$ 
and $f(x)$ in terms of TMDs $f(x, k_\perp)$. 

An important point is that the components of $k_\perp$ may take any 
values from $-\infty$ to $\infty$, even when the 
distribution in $k_\perp$ is mostly restricted to a limited range,
like in a Gaussian $e^{-k_\perp^2/\Lambda^2}$. 
Similarly, the $(y-x)P_3$ part of the $k_3$-distribution may take 
any values. As a result, $Q(y,P_3)$ formally has   the $-\infty <y <\infty$ support region,
though  possibly with a rapid decrease  (say, like $e^{-y^2 P_3^2/\Lambda^2}$) for large $y$.  

In other words, for a finite $P_3$, there is no requirement that the fraction $y$ 
is smaller than 1 or positive.
Even in a fast-moving hadron, there is some probability
that a parton moves in the opposite direction,
and hence, that some other parton has the momentum $k_3$ 
larger than $P_3$.  Still, with increasing $P_3$, 
the chances for fractions outside the $[0,1]$ segment decrease rapidly,
reflecting the large-$k_\perp$ dependence of the relevant TMD $f(x,k_\perp)$. 

 When $Q(y,P_3) \sim e^{-y^2 P_3^2/\Lambda^2}$, one may 
 consider $y^n$  moments of  quasi-PDFs $Q(y,P_3)$  calculated 
 over the whole \mbox{$-\infty <y<\infty$}  axis and study their   relation
 to  the $x^n$ moments of the light-cone PDFs $f(x)$. 
 
 Still, starting with   the first papers \cite{Ji:2013dva,Xiong:2013bka} on quasi-PDFs,  it was known
 that the simplest perturbative calculations produce 
 $\sim 1/|y|$ behavior for quasi-PDFs at  large $|y|$. 
 Such a behavior  reflects a  slow $\sim 1/k_\perp^2$  decrease 
 of the perturbative hard tail of TMDs in renormalizable theories. 
 Clearly, if  $Q(y,P_3) \sim 1/|y|$, then even the zeroth moment
 of $Q(y,P_3) $  diverges, so that it apparently makes no sense
 to consider $y^n$ moments of $Q(y,P_3) $. 
Since  the standard procedures of extracting PDFs from the lattice   
\cite{Lin:2014zya,Alexandrou:2015rja,Ma:2017pxb}
do not involve a calculation of the  moments,
the divergence of these moments did not attract much attention.

However, recently  it was argued by G.C. Rossi and M. Testa \cite{Rossi:2017muf,Rossi:2018zkn} 
that the divergence of the $y^n$ moments of $Q(y,P_3)$ 
poses a serious problem for   extraction of PDFs 
from lattice QCD simulations.  The basic  claim is that the 
infinite values of $\langle  y^n \rangle_Q$  quasi-PDF moments 
are  in conflict with  the finite values of the $\langle  x^n \rangle_f$
moments of the usual PDFs.

  Irrespectively of  these claims, we find that 
    the structure of quasi-PDFs $Q(y,P_3)$   outside 
  the central  $|y|\leq 1$  region is an interesting problem on its own,
  and we   analyze it in 
  the present paper.   Our study is based on   
  the concept \cite{Radyushkin:2017cyf} of the Ioffe-time pseudo-distributions (pseudo-ITDs) ${\cal M} (\nu, -z^2)$.
  They are basically the matrix elements $M(z,p)$  of bilocal operators
  $\sim \phi (0) \phi (z)$  treated as functions of the Lorentz 
  invariants, the Ioffe time  $\nu =-(zp)$ \cite{Ioffe:1969kf,Braun:1994jq} and  the invariant interval $z^2$. 
 [Our convention is to add  ``pseudo''  to the name of   distributions defined for nonzero $z^2$,
  and skip it for their light-cone analogs.]
  
  While ${\cal M} (\nu, -z^2)$  does not involve 
  momentum fraction variables like $y$ and $x$,
  quasi-PDFs $Q(y,P_3)$ and pseudo-PDFs ${\cal P} (x, z_3^2)$ 
  may be obtained  \cite{Radyushkin:2017cyf}   from ${\cal M} (\nu, -z^2)$ as  Fourier transforms. 
  The advantage of this approach is a direct use of the coordinate representation
  that greatly simplifies further  considerations of  pseudo-PDFs, TMDs  and quasi-PDFs.
  
  Furthermore, as we will show, the fact that the quasi-PDFs $Q(y, P_3)$ 
  do not vanish outside the $|y| \leq 1$ region, is directly connected 
  with the presence of  a non-trivial $z_3^2$- dependence   in the relevant pseudo-PDFs
   ${\cal P} (x, z_3^2)$. 
  
 The paper is organized as follows.  
In  Section  2,   we start with   reminding the definition   of the pseudo-ITDs and their relation to 
 pseudo-PDFs, 
quasi-PDFs and TMDs.   We write a formal $1/P_3^{2l}$ series expansion  for the 
$\langle y^n \rangle_Q$  moments of the quasi-PDFs in terms of the combined 
   $\langle  x^{n-2l} 
    k_\perp^{2l}  \rangle_{\cal F} $  moments of TMDs  ${\cal F} (x, k_\perp^{2})$.
    In the case of ``very soft'' TMDs, i.e., those   vanishing  faster than any inverse 
    power of  $k_\perp^2$  for large $k_\perp$, this expansion allows to study $\langle y^n \rangle_Q$
    moments (which are finite in this case) and their relation to $\langle x^n \rangle_f$
    moments of the usual PDFs. 
    
 In Section 3,  we study the consequences of having a hard $\sim 1/k_\perp^2$ tail
 of TMDs, present in renormalizable theories, including QCD. 
 In this  case, the combined  $\langle  x^{n-2l} 
    k_\perp^{2l}  \rangle_{\cal F} $  moments diverge.  For $l=0$, one has a 
    logarithmic divergence corresponding to the usual perturbative evolution.
    For $l \geq 1$, one faces power divergences equivalent to those discussed in Refs. 
    \cite{Rossi:2017muf,Rossi:2018zkn}.
    We show that they  reflect the  slowly $\sim 1/|y|$ decreasing perturbative contributions 
    to $Q(y,P_3)$.  We also show that the $|y|>1$ parts of $Q(y,P_3)$ are generated
    by the $z_3^2$-dependence of the pseudo-PDFs ${\cal P} (x, z_3^2)$. In Section 4, we study 
    possible forms of the $z_3^2$-dependence. 

In Section 5,  we discuss the  matching relations  connecting  the lightcone PDFs to  pseudo-ITDs
and quasi-PDFs.  According to  the operator product expansion (OPE),
the reduced \mbox{pseudo-ITD}   $\mathfrak{M} (\nu, z_3^2) $ is given by the $\overline{\rm MS}$-ITD ${\cal I} (\nu, \mu^2)$ 
plus ${\cal O} (\alpha_s)$ perturbative contribution that contains the $\sim \ln z_3^2$ term
responsible for the slowly varying $\sim 1/|y|$ terms in the $|y|>1$ part of the quasi-PDF $Q(y,P_3)$.
The latter, hence,  is given by the  $\overline{\rm MS}$-PDF  $f (x, \mu^2)$
plus  ${\cal O} (\alpha_s)$ perturbative contribution that contains the 
 slowly varying $\sim 1/|y|$ terms in the $|y|>1$ part.  
 
{\it Vice versa},  $f (x, \mu^2)$   is given by   the
 difference between the  lattice quasi-PDF $Q_L(y,P_3)$
 and  that  ${\cal O} (\alpha_s)$ perturbatively calculable  contribution.
 This means that the implementation of the matching condition includes a  subtraction,
 though not of the kind discussed by Rossi and Testa in Refs. \cite{Rossi:2017muf,Rossi:2018zkn}. 
  The final  point is that,  for large $P_3$, the quasi-PDF  
   $Q_L(y,P_3)$   must be purely perturbative in the $|y|>1$ region.
  Hence, the above difference vanishes outside the $|y|\leq 1$ segment,
  and the moments of the  light-cone PDF $f (x, \mu^2)$  extracted
  in this way are finite.
  
Section 6 contains summary  and conclusions.

 \setcounter{equation}{0}   


\section{Parton distributions} 

 \subsection{Ioffe-time distributions and  pseudo-PDFs}

 Defining a parton distribution either in a continuum theory or on the lattice, 
 one starts with 
 a  matrix element   $\langle  p | \phi (0)  \phi(z)  | p \rangle  \equiv M(z,p)$
 of  a product of two parton fields. 
  We 
 use  here simplified scalar notations,
since  the
 details of  parton
 spin structure  are not central to the concept
 of parton distributions, and may be added, if needed,  at later stages.

 By    Lorentz invariance,  $M(z,p)$ 
 is a  function of 
 two scalars,
 the {\it Ioffe time}
 \cite{Ioffe:1969kf,Braun:1994jq}   \mbox{$(pz) \equiv -  \nu$}  
 and   the \mbox{interval $z^2$}
 \begin{align}
 M(z,p)
 =  & {\cal M} (-(pz), -z^2)  \  . 
 \label{lorentz}
 \end{align}  
 
 As shown  in Refs. \cite{Radyushkin:2016hsy,Radyushkin:1983wh},
 for any  contributing  Feynman diagram,   
 the   Fourier transform of   $ {\cal M} (\nu, -z^2) $ with respect to 
 the Ioffe time $\nu$ 
 has the  $-1 \leq x \leq 1$ support, familiar from the studies of the usual parton densities, 
 \begin{align}
 {\cal M} (\nu, -z^2) 
 &   = 
 \int_{-1}^1 dx 
 \, e^{i x \nu } \,  {\cal P} (x, -z^2)  \  . 
 \label{MPD}
 \end{align}  

When $z$ is on the light cone, $z^2=0$, we deal with 
 the ordinary  (or light-cone) parton distributions 
\begin{align}
{\cal M} (\nu , 0)  =
\int_{-1}^1 dx \, e^{ix\nu} \,  f(x) \, 
 . 
\label{twist2}
\end{align}
 Thus, ${\cal P} (x, 0)=f(x)$, and  the 
 function  ${\cal P} (x, -z^2)$  generalizes
 the concept of PDFs 
 onto the case of non-lightlike \mbox{intervals $z$.}  Following \mbox{Ref. \cite{Radyushkin:2017cyf},} 
 we will refer to it as 
 {\it pseudo-PDF}  or parton {\it pseudo-distribution}  function.

\subsection{Quasi-PDFs}

The simplest example of a  spacelike  interval is obtained 
when just one component is nonzero,  $z= \{0,0,0,z_3 \}$. 
 Choosing  $p= (E, {\bf 0}_\perp, P)$,
one can  define the quasi-PDF  
\cite{Ji:2013dva}  as the Fourier transform of $M(z_3, P) $  with respect to $z_3$  
\begin{align}
Q (y, P) 
&   = \frac{P}{2\pi} \, 
\int_{-\infty}^\infty  dz_3 
\, e^{-i yPz_3 } \,  M(z_3, P)  
\label{Qdef} .
\end{align}   

	Combining  Eqs.  (\ref{MPD}) and (\ref{Qdef})  gives  a relation between 
	the quasi-PDF $Q(y,P)$ and the pseudo-PDF ${\cal P} (x,z_3^2)$
	corresponding to the $z=z_3$ separation
	\begin{align}
	Q(y,P) =\frac{P}{2 \pi}  \int_{-1}^1 dx \,   \int_{-\infty}^\infty dz_3\,
	e^{-i(y-x) Pz_3}\, 
	{\cal   P} (x, z_3^2)  \ . 
	\label{Py}
	\end{align} 
	One can see that though  the pseudo-PDFs have
	the $-1 \leq x \leq 1$ support,  the quasi-PDFs $Q(y,P)$   are  
	defined   
	for all    real $y$.  

Another observation is that if the pseudo-PDF  does not depend on $z_3^2 $, i.e.,	if 
${\cal   P} (x, z_3^2) =f(x)$, then the quasi-PDF $Q(y,P)$ does not depend on $P$,
and $Q(y,P)=f(y)$.  

Thus, it is the dependence of ${\cal   P} (x, z^2)$ 
(or, equivalently, of  ${\cal   M} (\nu, z^2)$) on $z^2$  that  determines the deviation 
of quasi-PDFs  from PDFs. In particular, it generates the parts of $Q(y,P)$ outside 
the PDF support region 
$|y|\leq 1$. 

    In    
QCD and other renormalizable theories,  the presence of 
the $z_3^2$-dependence    is  unavoidable, because   $ {\cal M} (\nu, z^2)  $ 
has  \mbox{ $\sim \ln z^2$}   contributions for small $z^2$.  Furthermore, 
these terms are  singular  in the  $z^2 \to 0$ 
limit which complicates the definition of  the light-cone PDFs.

\subsection{Transverse Momentum Dependent  PDFs}

A very convenient way to parametrize the $z^2$-dependence for a space-like $z$  is 
provided by a description in terms of  the transverse momentum dependent PDFs.
Using  again  $p= (E, {\bf 0}_\perp, P)$ and 
 choosing 
$z$ that has   only  $z_-$  and 
$z_\perp= \{z_1,z_2\}$  components, while    $z_+=0$, 
we have $\nu = - p_+ z_-$, 
$z^2=-z_\perp^2$, and 
the  TMD  is  defined by 
     \begin{align}
 {\cal P} (x,  z_\perp^2) 
&   = 
    \int  {d^2 k_\perp }      \,  e^{-i( k_\perp \cdot  z_\perp)} 
      {\cal F} (x, k_\perp^2)  \   . 
      \label{PTMD} 
\end{align}    
  Due to the rotational 
  invariance,
  this   TMD  depends on $k_\perp^2$  only. 
Integrating over the angle between ${\bf k}_\perp$ and $ {\bf z}_\perp$ 
 gives 
 \begin{align}
{\cal P} (x,  z_\perp^2) & = 
2 \pi \,   \int_{0}^{\infty} \, d k_\perp\, k_\perp    %
  J_0 \left ( k_\perp z_\perp \right )   \,  {\cal F } (x, k_\perp^2 )  \  , 
 \label{PcalF}
\end{align} 
where $J_0 $ is the Bessel function. 

Now recall 
   that ${\cal P} (x, -z^2)$ is a function  defined  in a covariant way 
 by Eq. (\ref{MPD}).
This implies  that this {\it TMD representation}   \cite{Radyushkin:2017ffo}  may be written   for a general \mbox{spacelike
$z$.}  One should just  change  $z_\perp \to \sqrt{-z^2}$ and  $k_\perp \to k$ in Eq. (\ref{PcalF}).  
In particular, one may take $z=\{0,0,0,z_3\}$, i.e., choose $z$ 
 in the  purely longitudinal direction,  and write 
 \begin{align}
{\cal P} (x,  z_3^2) = 
2 \pi \,   \int_{0}^{\infty}  d k\, k\,     %
  J_0 \left ( k z_3  \right )   \,  {\cal F } (x, k^2 )
   \  . 
 \label{McalF3}
\end{align}  
  While $ {\cal F } (x, k^2 )$
is a  function  that coincides with  the TMD, one does not
need   to specify a ``transverse'' plane  and treat $k$ as the magnitude
of a 2-dimensional momentum in that  plane.

  \subsection{Support mismatch} 

Using the TMD parametrization (\ref{McalF3})  in the quasi/pseudo-PDF relation 
(\ref{Py}), and expanding $  J_0 \left ( k z_3  \right ) $
into the Taylor series, 
we get  a   formal $1/P^{2l}$ expansion  for  the quasi-PDF $Q(y,P)$   
  \begin{align}
 Q(y, P) =  &     \sum_{l=0}^\infty  \int  d^2 k_\perp  
  \frac{k_\perp^{2l} } {4^l P^{2l} (l!)^2}     \,  
 \frac{\partial^{2l}}{\partial y^{2l}}
 {\cal F } (y, k_\perp ^2)
  \ . 
 \label{QyPDel}
\end{align} 
 To shorten formulas,  we have  switched here  back \mbox{$k \to k_\perp$}  in the  {\it  notation } 
 for the integration variable of the TMD representation
  (\ref{McalF3}), and also wrote the resulting $2\pi k_\perp dk_\perp$ as $d^2 k_\perp$. 
  We can  do this because the TMD ${\cal F} (x, k_\perp^2)$ does not depend on angles. 
  As a matter of  caution, we repeat again that $k$ or  $k_\perp$ should be understood 
  simply as scalar  variables of the TMD parametrization. There is no need to specify 
  in which plane $k_\perp$ is.

According to Eq. (\ref{Py}),  the quasi-PDF $Q(y,P)$  has the  \mbox{$-\infty <y<\infty$}  
support region.  However, the quasi-PDF $Q(y,P)$   in  Eq. (\ref{QyPDel}) 
is given by a sum  of terms involving the TMD $ {\cal F } (y, k_\perp ^2)$  that has the $-1\leq y \leq 1$ support. 
The explanation of  the  apparent discrepancy  is that the  innocently-looking 
derivatives of  ${\cal F} (y, k_\perp^2)$  in the expansion (\ref{QyPDel}) 
may generate  
an infinite tower of 
singular functions  like 
$\delta (y)$, $\delta (y\pm 1)$ and their derivatives. 
To this end, we recollect  that, even when a function $f(y)$ 
has a nontrivial support $\Omega$  (say, $-1\leq y \leq 1$), one may  formally represent 
it by a series 
 \begin{align}
 f(y) = \sum_{N=0}^\infty \frac{(-1)^N}{N!} \, M_N\, \delta^{(N)}(y) 
 \label{sumdel}
\end{align}
over the functions $\delta^{(N)}(y) $ with an apparent support at one point $y=0$ only.
Here, $M_N$ are the moments of $f(y)$,
 \begin{align}
  M_N\, = \int_{\Omega} dy \, y^N\,  f(y) \ . 
  \label{moments} 
\end{align}

Hence, the support mismatch may be explained by   the fact that the delta-function 
and its derivatives 
are 
integration  prescriptions (mathematical distributions) rather than ordinary  functions.
But  this also means that while  the difference between $Q(y,P)$ 
and $f(y)$ is formally given by a  series in powers of $1/P^{2}$,
its  coefficients are not the ordinary functions of $y$.

 \subsection{Moments of very soft  quasi-PDFs}

In order to get relations involving usual functions,
one may wish to integrate the equations  in which these  distributions enter,
e.g., to  take moments.  
Indeed, the derivatives disappear if  we calculate 
  the $y^n$ moments  $\langle y^n \rangle_Q$  of the quasi-PDFs
    \begin{align}
\langle y^n \rangle_Q \equiv   &  \int_{-\infty}^\infty dy \, y^n Q (y, P)  \nn & =
 \sum_{l=0}^{[n/2]} \frac{n!}{(n-2l)!(l!)^2}   \frac{ \langle  x^{n-2l} 
    k_\perp^{2l}  \rangle_{\cal F}  }{4^l P^{2l} } 
  \  , 
 \label{Qyn}
 \end{align} 
 where $\langle  x^{n-2l} 
    k_\perp^{2l}  \rangle_{\cal F} $ are the combined moments of TMDs
      \begin{align}
\langle  x^{n-2l} 
    k_\perp^{2l}  \rangle_{\cal F}  \equiv   &    \int_{-1}^1 dx\, x^{n-2l} 
      \int d^2 k_\perp \, k_\perp^{2l}  \, {\cal F} (x, k_\perp^2)   
  \  . 
 \label{combmom}
 \end{align}

    In the case of {\it very soft  distributions}  which vanish faster than any power of $1/k_\perp^2$ for large $k_\perp$,
    all the combined moments 
    $\langle  x^{n-2l} 
    k_\perp^{2l}  \rangle_{\cal F} $ are finite and Eq. (\ref{Qyn}) tells us that   then  
   $ \langle y^n \rangle_Q$  differs from    $\langle x^n \rangle_f$ by terms having the   
    $( \langle k_\perp^2 \rangle_{\cal F}/P^2)^l$ structure. 
    
 Two lowest moments $n=0$ and $n=1$ do not  involve $l\geq 1$  terms.
 For the normalization integral, \mbox{Eq. (\ref{Qyn})}  gives 
  \begin{align}
 \int_{-\infty}^\infty dy \, Q(y, P)  = & \, 
 \int_{-1}^1 dx\,   \int d^2 k_\perp \,  {\cal F} (x, k_\perp^2)  %
\,  \nonumber   \\ &
 =    \int_{-1}^1 dx\, f(x)
   \ . 
 \label{Qy0}
\end{align}  
Thus, the area under $Q(y,P)$  does not change with $P$ and is equal to the area under $f(x)$,
the phenomenon corresponding  to the quark number  conservation.

Similarly,  the first $y$-moment is  given by 
  \begin{align}
 \int_{-\infty}^\infty dy \,y\,  Q(y, P)  = &      \int_{-1}^1 dx\,  x \, f(x) 
   \ ,
 \label{Qy1}
\end{align} 
which corresponds to the momentum conservation. These two sum rules 
have been originally derived in our paper \cite{Radyushkin:2016hsy}.

 \setcounter{equation}{0}   
     
    \section{Hard part} 
 
 \subsection{Perturbative evolution}

    
          In renormalizable theories (most importantly, in QCD, but also in models with Yukawa gluons), 
  i.e.,  theories having a dimensionless coupling constant $g$,
  the perturbative  corrections to all  ``twist-2''  $\phi (0) \phi (z)$-type correlators 
  (in QCD   we have in mind $\bar \psi (0) \Gamma \psi (z)$  
   quark 
  and $G(0) G(z)$  gluon operators) unavoidably 
  contain terms that are logarithmic in $z^2$ for small $z^2$, e.g. $\sim g^2 \ln (-z^2 m^2)$ at 
  one-loop level, $m$ being some infrared cut-off. 
  For DIS structure functions $F(x_{\rm Bj}, Q^2)$,  such terms produce  the logarithms 
  $\sim g^2 \ln (Q^2 /m^2)$
  generating their  perturbative evolution 
   \cite{Altarelli:1977zs,Gribov:1972ri,Dokshitzer:1977sg}
  with $Q^2$.
  
  For pseudo-PDFs $ {\cal P} (x,  z_\perp^2) $
  that define TMDs through \mbox{Eq. 
  (\ref{PTMD}), } 
  the  \mbox{$\sim g^2 \ln (-z^2 m^2)$}   terms  result in the $\sim g^2 \ln (z_\perp^2 m^2)$ contributions 
  for small $z_\perp$.  The 2-dimensional Fourier transform with respect to $z_{\perp}$
  converts such terms into contributions with a $\sim 1/k_\perp^2$ ``hard tail'' 
  for large $k_\perp$ (see, e.g., Ref. 
   \cite{Radyushkin:2016hsy}). 
   
  Thus,  in general, 
  TMDs  ${\cal F} (x, k_\perp^2)$ in renormalizable theories {\it must} have 
     a  {hard part} 
    that has the $1/k_\perp^2$ behavior for large $k_\perp$.
 For non-singlet densities in QCD,  it  is 
    given at one loop by 
     \begin{align} 
 {\cal F}^{\rm hard}  (x, k_\perp^2)   = \frac{ \Delta (x) }{\pi k_\perp^2} \  , 
 \label{Fhard} 
   \end{align}   
 where   $\Delta (x) $ is  obtained from the  PDF  $f^{\rm soft} (x)$ 
 (corresponding to  a primordial  soft TMD)  through       
   \begin{align}
\Delta (x)  =  \frac{\alpha_s}{2\pi} \, C_F\,  \int_{x}^1 \frac{du }{  u } B(u) 
f^{\rm soft} (x/u) 
\label{Delta-x} \  ,  
\end{align}
and  $ B(u) $  is 
 the 
 Altarelli-Parisi  (AP) evolution kernel  \cite{Altarelli:1977zs} 
 \begin{align} 
B(u)  =&
       \left [\frac{1+u^2} {1-u}   \right ]_+
  \ . 
     \label{V1}
  \end{align} 
  Since the parton densities 
  $f(x,\mu^2)$ are obtained from the TMDs by a 
  $d^2 k_\perp$ integration, the well-known 
   logarithmic evolution of $f(x,\mu^2)$  
   with a cut-off $\mu$, is a direct consequence of 
   the $1/k_\perp^2$ behavior of  the relevant TMDs in QCD.
  
    If one calculates the combined 
    moments      $\langle  x^{n-2l} 
    k_\perp^{2l}  \rangle_{\cal F} $     for the hard   term, they  diverge, starting from the lowest $l=0$ moment
    in $k_\perp^2$.  
    In the $l=0$ case, the divergence is  logarithmic.
    Let us see that 
     it  just reflects the fact  that the quasi-PDF  $Q(y,P)$ 
    for large $P$ in this case  has the logarithmic  perturbative evolution with respect to $P^2$.
 To begin with, we write  the hard part   in  the coordinate representation  
     \begin{align}
 {\cal  P}^{\rm hard}  (x, z_3^2) =&   \,-      \ln  (z_3^2   m^2)  
\Delta (x)  \  ,
 \label{hardP}
\end{align} 
where $m$ is some infrared regularization scale.
Rewriting  the quasi-PDF  definition  in terms of  the
pseudo-ITD as  
\begin{align} 
Q(y,  P)   =\frac{1}{2 \pi}  \int_{-\infty}^{\infty}  d\nu \, 
\, e^{-i y  \nu}  \, {\cal M} (\nu,  \nu^2/P^2)    \   
\label{QMnn}
\end{align}  
we find  that 
   \begin{align}
&  {\cal  M}^{\rm hard}  (\nu, \nu^2/P^2) =   \,-    \frac{\alpha_s}{2\pi} \, C_F\,  \ln  (\nu^2 m^2/P^2 )  
\nn & \times 
\int_0^1  du \,  B (u) \   \int_{-1}^1 dx 
 \, e^{-i u x \nu } \, f^{\rm soft} (x)   \  .
 \label{hardP}
\end{align} 
  As a result, the hard part of the quasi-PDF $Q(y,P)$  has the evolution  $\ln P^2$ part
   \begin{align}
 Q^{\rm ev}  (y, P) & =\ln  ({P^2/m^2} )  \, \Delta (y)  \ . 
 \label{Qlog}
\end{align} 

Comparing with Eq. (\ref{Fhard}), we conclude that,
  calculating the evolution part, one should  
cut-off the $k_\perp$ integral 
at  $|k_\perp |\sim P$ values, so that 
 it  is given by 
 \begin{align}
 Q^{\rm ev}  (y, P) & =\int_{|k_\perp |\lesssim P} 
   d^2 k_\perp\,    {\cal F}^{\rm hard}  (y, k_\perp^2) \simeq \ln (P^2) \, \Delta (y) \  . 
 \label{QlogF}
\end{align}

     \subsection{Two lowest moments} 
     \label{lowmom}

  As we have seen, for very soft distributions,  the $n=0$ and \mbox{$n=1$}   moments of  quasi-PDF $Q(y,P)$
     coincide with these    moments of the PDF $f(x)$. To proceed with the hard part, 
     we   use  
   \begin{align}
   &
\int_0^1 dx \, x^n \, \Delta (x)  = -  \frac{\alpha_s}{2\pi} \, C_F\, \gamma_n 
 \int_0^1 d\zeta \,  \zeta^n \,
f^{\rm soft} (\zeta)  \  , 
\label{Delta-xn}
\end{align}
  where $\gamma_{n}$'s are   related to anomalous dimensions 
of operators with $n$ derivatives, 
  \begin{align}
   &
\gamma_n =  - \int_0^1 du \,  u^n \,
B(u) \ . 
\label{gamman}
\end{align}
Thus, 
for the zeroth moment  of $ Q^{\rm ev}  (y, P) $,  the coefficient 
in front of   $\ln P^2 $  is proportional to
 the anomalous dimension $\gamma_0$ 
of the vector current. Since   $\gamma_0$
 vanishes, the area under $Q(y,P)$  does not change with $P$ and is equal to the area under $f(x)$,
the phenomenon corresponding  to the quark number  conservation.

Similarly,  the first $y$-moment  of the hard part   of $Q(y,P)$
has the $\ln P^2$ part    proportional 
to the  anomalous dimension \mbox{$\gamma_1=4/3$}  that is nonzero.  This 
reflects  the fact that the quark-gluon interactions  change  the 
momentum carried by the quarks,
and only the total  momentum of quarks plus gluons is conserved in the evolution process.

 \subsection{Higher  moments  and large-$|y|$ behavior} 

  \label{highmom} 
  
According to the general formula (\ref{Qyn}), the $y^2$-moment is given by 
  \begin{align}
\langle y^2 \rangle_Q = \langle x^2 \rangle _{{\cal F}}  + \frac{   \langle k_\perp ^2 \rangle_{{\cal F}} }{ 2 P^2}
 \ , 
 \label{Qy2}
\end{align}  
(see also Ref. \cite{Broniowski:2017gfp}), where 
  \begin{align} 
   \langle k_\perp ^2 \rangle_{{\cal F}} 
   = \int_{-1}^1 dx\,       \int d^2 k_\perp \, k_\perp^{2}  \, {\cal F} (x, k_\perp^2)   \  . 
   \label{k2}
\end{align} 
  When   ${\cal F} (x, k_\perp^2)$ vanishes faster than $1/k_\perp^4$ for large $k_\perp$,  the \mbox{$k_\perp$-integra}l converges. 
  Then  the difference between $\langle y^2 \rangle_Q$  and $ \langle x^2 \rangle _{{\cal F}}$
    decreases as $  \langle k_\perp ^2 \rangle_{{\cal F}}/2P^2$ for large  $P$. \footnote{W. Broniowski  and   E. Ruiz-Arriola
    \cite{Broniowski:2017gfp} 
    have checked that quasi-PDFs obtained by ETMC  
\cite{Alexandrou:2016eyt} satisfy Eq. (\ref{Qy2}), with $ \langle k_\perp ^2 \rangle_{{\cal F}} = 0.27$ GeV$^2$.}

However,  for a hard $\sim 1/ k_\perp^2$  TMD,  the  $ \langle k_\perp ^2 \rangle_{{\cal F}}$ integral diverges quadratically.
If, by analogy with  Eq. (\ref{QlogF}),  we  would set  the upper limit of $k_\perp $ integration 
    to be proportional  to $P$, 
the \mbox{$k_\perp^2$-weighted}  integral (\ref{k2}) 
 would be 
     proportional to $P^2$.   
     
      Because of the   compensation
   of the initial $1/P^2$ suppression factor by the $P^2$ factor
   resulting from the quadratic  divergence of the $k_\perp$-integral,
     the   contribution of the 
     $ \langle k_\perp ^2 \rangle_{{\cal F}}/2P^2$   term   does not disappear in the $P\to \infty$ limit. 
     One may also argue that,  on  the lattice,  
     the upper limit on the $k_\perp$  integral may be  set by the  lattice spacing. 
    Then,   a  cut-off for the $k_\perp$ integral  at the  $\sim 1/a$ value would 
result  in a $\sim 1/a^2 P^2$ 
contribution.

These  worries have been  formulated in recent  papers by \mbox{G.C. Rossi} and M. Testa  \cite{Rossi:2017muf,Rossi:2018zkn},
 who warned that one might need to
perform a nonperturbative {\it subtraction}   of such terms in lattice calculations. 
 The questions raised in Ref. \cite{Rossi:2017muf} 
have been   subsequently   addressed in Ref.  \cite{Ji:2017rah} by X. Ji {\it et al.}, who stated that 
 the extraction
of PDFs does not involve taking  moments of  quasi-PDFs.
It was also argued that the moments of quasi-PDFs do not exist
because $Q(y,P)$  decreases as $1/|y|$ for large $y$. 
While we agree with these statements in general, 
we think that the problem deserves a more detailed investigation.

 	 \setcounter{equation}{0}

       \section{Sources of $z^2$ dependence}

    As we discussed already,  the \mbox{$|y|>1$}    parts of  quasi-PDFs $Q(y,P)$
    are generated by the \mbox{$z_3^2$-dependence}    of the ITD ${\cal M} (\nu, z_3^2)$.
	 In particular, for large $z_3^2$,
	${\cal M} (\nu, z_3^2)$  has a fast decrease with $z_3$.
	This reflects  a finite size of the system. 
	 Such a behavior 
	 should appear   in any reasonable theory/model used 
	 to describe hadrons. 
	 The second type of the $z_3^2$-dependence 
	appears  in renormalizable theories.  As already mentioned,  then  $ {\cal   P} (x, -z^2)$ and 
	$ {\cal M} (\nu,  -z^2) $  contain, for small $-z^2$,   the terms \mbox{$\sim \ln (-z^2)$}  
	corresponding to   the  $\sim 1/k_\perp^2$ hard tail of ${\cal  F} (x, k_\perp^2)$.
	The tail is  generated  by  hard  gluon exchanges and 
	is proportional  to a small parameter $\alpha_s/\pi \sim 0.1$.

	Finally,  in QCD (and other gauge theories),  there is the third source of the \mbox{$z^2$-dependence}  related
	to some  special   contributions  originating from   the gauge link.
	These contributions vanish on the light cone $z^2=0$,   but do not  vanish for 
	spacelike $z^2$. Moreover, they contain link-specific UV divergencies,
	similar to those one encounters in the heavy-quark effective theory (HQET).  	
	Let us discuss  these types of $z^2$-dependence. 
	
	\subsection{Long-distance $z^2$-dependence}

	To begin with,  $ {\cal   P} (x, z_3^2)$ describes a finite-size system 
	(moreover, a system of confined quarks). Hence,  it should rapidly decrease
	for large $z_3$, say, like a Gaussian $\sim e^{-z_3^2 /R^2}$ or an exponential $\sim e^{-z_3 /R}$, 
	 where $R$ characterizes  the size  of the system. 
	A finite size of the system imposes no restrictions    on the behavior  of  $ {\cal   P} (x, z_3^2)$ 
	for small $z_3^2$.  Such a behavior is determined 
	by the short-distance dynamics. In  models  involving just soft interactions,
	one would expect that 
		$ {\cal   P} (x, z_3^2)$ is finite   in the 
		$z_3 \to 0$ limit,  like in the Gaussian and exponential cases.  Then 
	one may simply take $z_3=0$  in $ {\cal   P} (x, z_3^2)$
	to get $f(x)$.   
	In terms of TMDs, soft models usually are chosen 
	to have  a Gaussian $e^{-k_\perp^2/\Lambda^2}$
	or a power-law $\sim 1/(k_\perp^2+\Lambda^2)^n$ behavior for large $k_\perp$. 
	If   $n>1$, then the relevant pseudo-PDFs are   finite for $z_3^2=0$.  

	\subsection{Evolution-related  $z_3^2$-dependence}
	
	Since the small-$z^2$ limit in QCD is perturbative,
	one would  expect that the only singularities of $ {\cal   P} (x, -z^2)$
	for $z^2=0$ are those generated by perturbative corrections.
	As  already mentioned, at one loop one gets $\sim \alpha_s \ln (-z^2)$ terms. 
	Hence,  it makes sense to treat 
	$ {\cal P}  (x, -z^2)  $    
	as a sum of a ``primordial''   soft part  $ {\cal P}^{\rm soft} (x, -z^2)$ that 
	 has a finite $z^2\to 0$ limit, 
	and a  logarithmically singular {\it hard part}      reflecting   the evolution,
	and generated by hard gluon corrections to the original purely soft function. 
	The same applies to   $ {\cal M}  (\nu, -z^2)  $.

	A singularity at $z^2=0$ means that  the lightcone  object ${\cal   M} (\nu, -z^2=0)$ is a divergent quantity.
	In perturbative calculations of the  lightcone matrix element, the $\ln (-z^2)$ singularities  convert  into ultraviolet 
	logarithmic divergences. These UV divergences are then additional  to the usual 
	UV divergences related to the propagator  and vertex 
	renormalization.  
	
	Still, as far as $z^2$ is kept finite,  one does not  have 
	these  additional  UV  divergences,
	and does not need to introduce a  regularization for the $\bar \psi (0) \ldots \psi (z)$ operator. 
	One should deal  with the usual UV divergences and their renormalization only. 
	Such a renormalization (characterized by some parameter $\lambda$) would produce (in a covariant gauge, say) 
	just a trivial $Z_\psi (\lambda/m)$ renormalization factor for the $\psi$-fields  ($m$ being  an infrared cut-off, e.g., a mass of the 
	$\psi$ field).  This factor is the same whether $z$ is on the light cone or not. 
	
	Except for this trivial 
	dependence on the UV cut-off 
	$\lambda$, 
  the pseudo-ITDs ${\cal   M} (\nu, -z^2)$   in a general renormalizable (but non-gauge) theory,
	  depend on $\nu$ and $z^2$ only.  The $\ln (-z^2)$ terms are just a particular 
	  form of the $z^2$-dependence, and they do not require any regularization as far as $z^2$ is finite,
	  which is the case in lattice simulations.

	Theoretically, one may take $z$  on the light cone. Then  one should  regularize the  resulting extra UV divergences
	in some way, e.g.,  by 
	imposing a momentum cut-off  or   by incorporating the $\overline{\rm MS}$ scheme, etc.
	The  resulting {\it lightcone} ITD ${\cal I} (\nu, \mu^2)$ 
	\begin{align}
{\cal I} (\nu, \mu^2) 
&   = 
\int_{-1}^1 dx 
\, e^{i x \nu } \, f   (x, \mu^2)  \ 
\label{ITD}
\end{align}   
  introduced in Ref. 
 \cite{Braun:1994jq}
	naturally  depends on the parameter  $\mu$  involved in the regularization
	of these ultraviolet divergences generated by taking $\ln z^2$ for $z^2=0$. 
		
	\subsection{UV singular terms generated by the gauge link}
	
	Furthermore, in QCD, the gauge link factor connecting  $\bar \psi (0)$ and  $\psi (z)$
	generates contributions that are absent on the light cone, and
	moreover,  are ultraviolet divergent. These divergences 
	may be regularized
	using, e.g.,  the Polyakov prescription \cite{Polyakov:1980ca} $1/z^2\to 1/(z^2-a^2)$  for the gluon propagator 
	in the coordinate space. Then  one finds that,
	for a fixed UV cut-off $a$, these terms vanish in the $z_3^2 \to 0$ limit,
	like $|z_3|/a$ for the linear UV divergence and like $\ln (1+z_3^2/a^2)$ for the logarithmic one.
	That is why such terms are invisible on the light cone. 
	Hence, we must make an effort to completely exclude these terms from   ${\cal   M} (\nu, z_3^2)$. 
We emphasize that we  need  to {\it eliminate}  the  terms invisible in the light-cone  limit even if they are UV finite.
	
	As a matter of fact, in QCD they are UV divergent, 
	and this fact has shifted the whole subject to the discussion of the UV divergences. 
	 These UV divergences 
		 were considered as  the main problem  in many recent papers
	  \cite{Ji:2017oey,Ishikawa:2017faj,Green:2017xeu,Izubuchi:2018srq}. 
	Having UV singularities,   one should add the regularization  parameter ($a$ in this case)
	to the argument of the regularized pseudo-ITD: ${\cal   M} (\nu, -z^2) \to {\cal   M} (\nu, -z^2;a)$.	
	These  UV divergences are similar to those 
	known  from the HQET studies,
	and are   multiplicatively renormalizable \cite{Ishikawa:2017faj,Ji:2017oey,Green:2017xeu}.
	
	Since the parameter $a$ appears only in the combination $z_3/a$,
	the UV-sensitive terms form a factor $Z(z_3^2/a^2)$.  As discussed above, this factor is an artifact  
	of having a non-lightlike $z$, and has nothing to do with the lightcone  PDFs.
	Thus, constructing the latter, we   should  {\it exclude}  $Z(z_3^2/a^2)$ 
	from the pseudo-ITD ${\cal   M} (\nu, z_3^2;a)$. 
	In other  words, one should build quasi-PDFs from the modified 
	function $Z^{-1}(z_3^2/a^2) {\cal   M} (\nu, z_3^2;a)$. 
	
	By construction, $Z^{-1}(z_3^2/a^2) {\cal   M} (\nu, z_3^2;a)$ does not have 
	\mbox{$a \to 0$}   UV divergences. 
	However, if  the goal  is just to  remove  the divergences,  then 
	one may use any  combination  of  the $Z^{-1}(1/\mu_{\rm UV}^2 a^2)
	{\cal   M} (\nu, z_3^2;a)$ type for the renormalized ITD.
	But the result  then will 
	have the dependence on the renormalization scale 
	$\mu_{\rm UV}$. The renormalized ITD will also contain 
	the $z_3^2$-dependence of the $Z(z_3^2/a^2)$-factor,
	that should be excluded in the construction 
	of the light-cone PDFs.   In the approaches of Refs. \cite{Ji:2017oey,Green:2017xeu,Izubuchi:2018srq},
	 this is done at the final stage, when the matching conditions are  applied. 
	
Our point of view is  that  it is more beneficial   to 
remove  the UV divergences    together with the associated
	\mbox{$z_3^2$-dependence}    from the very beginning.   This may be done  
	by multiplying ${\cal   M} (\nu, z_3^2;a)$ with  the 
	$Z^{-1}(z_3^2/a^2)$ factor.  To do this, one should know the 
	$Z(z_3^2/a^2)$ factor. 
Another possibility,  proposed 
 in our paper   \cite{Radyushkin:2017cyf},
is  to use the reduced pseudo-ITD
 \begin{align}
{\mathfrak M} (\nu, z_3^2;a) \equiv \frac{ {\cal M} (\nu, z_3^2;a)}{{\cal M} (0, z_3^2;a)} \  . 
 \label{redm0}
\end{align}
Then the   UV-sensitive factor $Z(z_3^2/a^2)$   automatically 
cancels  in the ratio (\ref{redm0}), since it is $\nu$-independent.  
So, there is no need to know it explicitly. 
The $Z_\psi (\lambda/m)$ factors   reflecting  the anomalous dimensions 
of the $\psi$ fields also cancel in the ratio (\ref{redm0}). 
The resulting function    has a finite  $a\to 0$ limit, which will be denoted
by ${\mathfrak M} (\nu, z_3^2) $.  This function {\it does not}   depend
on {\it any} UV  cut-off or  a  UV renormalization scale like $\mu_{\rm UV}$.

We may say that ${\mathfrak M} (\nu, z_3^2) $ is a {\it physical observable},
just like the deep inelastic (DIS) structure functions  $W(x_{\rm Bj}, Q^2)$. The latter 
depend on the external variables $x_{\rm Bj}$,  $Q^2$, but do not depend
on any ultraviolet cut-off or a  renormalization scale $\mu$, even if they are calculated in a 
renormalizable theory. 

 A widespread  statement is that $W(x_{\rm Bj}, Q^2)$
describes the hadron at  the distance scale $\sim 1/Q$. 
In this sense, ${\mathfrak M} (\nu, z_3^2) $ and the pseudo-PDF ${\cal P} (x,z_3^2)$,
by construction, 
 describe
a hadron at the  distance  $z_3$,  literally.

Thus, for the reduced ITD ${\mathfrak M} (\nu, z_3^2) $,
  there are  just two sources of the  $z_3^2$-dependence:
the long-distance nonperturbative dependence reflecting the finite size of the system,
and the short-distance perturbative $\sim \ln z_3^2$ dependence
related to  the usual perturbative  evolution. 
In this respect, the reduced pseudo-ITD ${\mathfrak M} (\nu, z_3^2) $ in QCD 
  has the $z_3^2$-structure similar to that in non-gauge renormalizable theories,
  in which we also have just two first types of the $z_3^2$-dependence. 
	
	 \setcounter{equation}{0}   
	 
\section{Matching}

The   
 relations  for the moments, like the formula  (\ref{Qy2}) for $\langle y^2 \rangle_Q$, and the 
general formula  (\ref{Qyn}), 
 that have been used
in our preceding discussion, 
are based on the Taylor expansion of    ${\cal P} (x,  z_3^2)$  over $z_3^2$. 
 Rossi and Testa in Refs. \cite{Rossi:2017muf,Rossi:2018zkn}  also  appeal to 
a  Taylor expansion in $z_3$. 
The basic reason for using the Taylor expansion is that the \mbox{$z_3$-dependence} 
of the matrix element is, in general,  unknown.  So,  a natural idea  is to  parametrize it  through the
values of the matrix elements of local operators. 

While this may be reasonable in a {\it very soft} case (in which all the derivatives with  respect 
to $z_3^2$ exist at $z_3^2=0$),  it  is clear that 
to use the  Taylor expansion at \mbox{$z_3^2=0$}   for the {\it hard}  logarithm $\ln z_3^2$
is problemetic. 
Fortunately,  the hard contribution  also has an advantage:  its \mbox{$z_3^2$-dependence}  at small $z_3^2$ 
(unlike that of the soft contribution) is known: at one loop it is given by $\ln  z_3^2$. 
Thus, if one needs to find  a quasi-PDF corresponding to
the $\ln z_3^2$ part of the matrix element, one  can do this  by simply calculating
the Fourier transform   of   $\ln z_3^2$  dictated by the quasi-PDF definition  
 (\ref{Qdef}) rather than to  use a Taylor expansion at a singular point. 
 
 \subsection{OPE and matching conditions for ITDs} 

When $\ln (-z^2)$ terms are present, a formal  light-cone limit $z^2 \to 0$ is  singular.
 Still, the  PDF community 
 wants   lattice predictions for  {\it the light cone} PDFs. 
In the continuum, the singular nature of the $z^2 \to 0$ limit  is 
 perceived  as an ultraviolet divergence in the  Feynman integrals for   operators on the light cone.
 It is worth repeating once more that these UV divergences 
 are just a consequence of our desire to take $z^2=0$.
 As far as $z^2$ is finite, these divergences are absent.

 To work  at  $z^2=0$, we need to  arrange 
an UV cut-off  for these hand-made divergences. 
Using,  say, the dimensional regularization and $\overline{\rm MS}$ scheme,
one would define  the light-cone ITD (\ref{ITD})  ${\cal I} (\nu, \mu^2)$. Its connection to    the pseudo-ITD
${\mathfrak M} (\nu, z_3^2)$ is given by the operator product expansion.
At one loop in QCD,  we have  \cite{Ji:2017rah,Izubuchi:2018srq,Radyushkin:2017lvu,Radyushkin:2018cvn}
  \begin{align} 
  {\mathfrak  M} ( \nu, z_3^2)   = &{\cal I}  (\nu, \mu^2)  -  \frac{\alpha_s}{2\pi} \, C_F\,  
\int_0^1  du \,    {\cal I}  (u \nu,\mu^2)    \nn &
 \times  \left  \{   B(u) \, \left [ \ln \left (z_3^2\mu^2 \frac{  e^{2\gamma_E}}{4} \right )  +1 
 \right ]   \right. \nn & \left. 
+   \left [ 4  \frac{\ln (1-u)}{1-u} - 2 (1-u)  \right ]_+ \right  \} +{\cal O} (z_3^2)
\   . 
\label{MNL0}
 \end{align}

 The OPE tells us that, for small $z_3^2$, the dependence of 
 $ {\mathfrak  M}  ( \nu, z_3^2) $   on $z_3^2$ {\it must} be given 
 by the $\ln z_3^2$ term on the right-hand side.
Hence, to get the light-cone ITD  ${\cal I}  (\nu, \mu^2) $ from, say,  
lattice calculations of $ {\mathfrak  M}  ( \nu, z_3^2) $, one should {\it subtract}
from the lattice pseudo-ITD $ {\mathfrak  M}  ( \nu, z_3^2) $ its   perturbative  $\ln z_3^2$  part
present in the r.h.s. of Eq. (\ref{MNL0}).
For an appropriately chosen/fitted $\alpha_s$, the result of such  a 
subtraction should be $z_3^2$-independent. 
 Such a procedure of extracting ${\cal I}  (\nu, \mu^2) $ from the lattice data 
 of \mbox{Ref. \cite{Orginos:2017kos}}  was described in our Ref.  \cite{Radyushkin:2018cvn}.

\subsection{Matching conditions for quasi-PDFs} 

Multiplying  Eq. (\ref{MNL0})  by $P e^{-i y z_3 P}$ and integrating
over $z_3$, we get  a  relation between the 
quasi-PDF $Q(y,P)$   (obtained from the reduced pseudo-ITD)  and the  light cone PDF $f(x,\mu^2)$. It  
has  the following structure 
  \begin{align} 
Q(y,P)  = &   f   (y, \mu^2)     -  \frac{\alpha_s}{2\pi} \, C_F\,  
\int_0^1  \frac{du}{u}  \,   f (y/u,\mu^2)   \nn &
 \times  \left  \{   B(u) \,  \ln \left (\mu^2/P^2 \right )  + C(u)  
   \right  \}   
   \nn &  +  \frac{\alpha_s}{2\pi} \, C_F\,  
\int_{-1}^1  {dx} \,   f (x,\mu^2)  \, L(y, x) +{\cal O} (1/P^2)
\   , 
\label{fNL}
 \end{align}
where the kernel $L(y,x;P)$ is formally given by
	\begin{align}
	L(y,x) =& -  \frac{P}{2 \pi}   \int_{0}^1 \, du \, B(u)  \nn & \times 
	 \int_{-\infty}^\infty dz_3\, 
	e^{-i (y-ux)z_3 P}\, \ln (z_3^2 \, P^2) \,  
	 \ . 
	\label{Lyx}
	\end{align} 
It involves the Fourier transform of $\ln z_3^2 $ and,  for large $P$,   it is   \mbox{\it the  only} 
perturbative  term that  produces 
contributions in the $|y|>1$ region. 
Eq. (\ref{fNL}) tells us that  the quasi-PDF $Q(y,P)$ {\it must} 
have  ${\cal O}(\alpha_s)$  contributions 
in the  \mbox{$|y|>1$}  region. 
In actual lattice calculations it is desirable (though challenging)  to try to check if the  lattice quasi-PDF  
 in the \mbox{$|y|>1$}  region is indeed close to  the convolution of the  fitted PDF with the $L$-kernel.

For large $P$, the soft contributions disappear from the 
$|y|>1$ region, and  the perturbative terms are  the only ones 
remaining  for  $|y|>1$.   This means that extracting  the PDF $  f   (y, \mu^2)  $
from the lattice data for $Q(y,P)$,  one deals with the combination,
 the 
``reduced'' quasi-PDF 
  \begin{align} 
  \widetilde Q(y,P) \equiv 
Q(y,P)  -   \frac{\alpha_s}{2\pi} \, C_F\,  
\int_{-1}^1  {dx} \,   f (x,\mu^2)  \, L(y, x) 
\   , 
\label{QNL}
 \end{align}
that   vanishes  in the $|y|>1$ region for large $P$ 
(provided that  we trust perturbative QCD!).
We may say that 
the $f \otimes L$   contribution {\it  
cancels} the perturbative slow-decreasing terms of the \mbox{$|y|>1$}  part  of $Q(y,P)$.
After   that, all the  remaining terms  in Eq. (\ref{fNL}) have the $|y|\leq 1$ support. 

In other words,  the process of getting  $\overline{\rm MS}$ PDFs  from quasi-PDFs
involves a {\it subtraction}  of  the  perturbative $|y|>1$ contributions  generated by the $\ln z_3^2$  term.

\subsection{Hard part of quasi-PDFs}  
 
An evident
observation from 
the study of  the hard contribution is  
     that the quasi-PDFs do not simply convert into the usual PDFs in the large-$P$ limit.
     They convert into PDFs only in the case of  {\it  soft}  TMDs and quasi-PDFs generated from them.
     
    When  the   hard part is included,  $Q(y,P)$ contains  the  terms  that are not present in 
      the lightcone PDFs and  which  are, moreover,     finite (for a fixed  $\alpha_s$) 
       in the $P \to \infty $ limit.
     Such terms appear  both in the ``canonical'' \mbox{$-1\leq y \leq 1$}  region and,  most importantly, 
outside it. 
     The presence of such terms was known since  the  first papers on quasi-PDFs \cite{Ji:2013dva,Xiong:2013bka}. 
  
In the context of pseudo-PDFs, these terms are generated by  the Fourier transform   
of  the $\ln z_3^2 $  hard term. 
In the momentum representation, $\ln z_3^2 $ (equivalent to   $\ln z_\perp^2 $)  corresponds to the  $1/k_\perp^2$ behavior, which needs some 
infrared  regularization. Let us choose the  mass-type  modification  \mbox{$1/k_\perp^2 \to 1/(k_\perp^2+m^2)$}.
Then $  \ln  (z_3^2)   \to- 2 K_0 (z_3 m)$, and we have (see Ref.  \cite{Radyushkin:2017lvu}) 
 \begin{align}
  Q^{\rm hard}  (y, P)  & =
 C_F \, \frac{\alpha_s}{2 \pi}  \, \int_{-1}^1 \, 
 \frac{dx} {|x|}  R(y/x, m^2/ x^2P^2)
 \  f^{\rm soft} (x) 
 \  , 
 \
 \label{QR}
\end{align} 
where the kernel $R(\eta, m^2/ P^2) $ \, 
  is given by 
   \begin{align}
 R(\eta;m^2/ P^2)  &= \int_{0}^1 du \,  \frac{ B(u) }{\sqrt{(\eta  -u )^2+m^2/ P^2}} 
  \  .
 \label{Reta}
\end{align}

 In lattice extractions, the real part of the pseudo-ITD corresponds to an even function of $y$,
 while the imaginary part corresponds to an odd function of $y$.
 Hence, in both cases, it is sufficient to consider   positive $y$ only. 
  For $\eta$, we  need  then to analyze   three regions, $\eta<0$, $0\leq \eta \leq 1$ and $\eta>1$. 
  
  In the central $0\leq \eta \leq 1$ region,  the $P\to \infty $ limit is singular,   reflecting 
  the presence of the evolution $\sim \ln P^2/m^2$ term  (\ref{Qlog}).
   There are also   terms  \cite{Radyushkin:2017lvu}
    \begin{align}
  R^{\rm middle } & (\eta) =
 \frac{1+\eta^2}{1-\eta} \ln \left [4 \eta  (1-\eta)  
   \right]
 \nn & 
+\frac{3/2}{ 1-\eta} +4\frac{\ln (1-\eta)}{1-\eta} - 1+2 \eta \   
\label{central}
   \end{align} 
 that   are  independent of $P$ in the $P\to \infty $
  limit. 
  For $|y|>1$, we can neglect $m^2/P^2$ 
in the $P\to \infty $ limit and get 
  \begin{align}
  Q^{\rm hard, out}  (y, P\to \infty)  =   &  \frac{\alpha_s}{2\pi} \, C_F\, 
 \int_{0}^1 \frac{dx}{x}  \,R (y/x;0) \,  f^{\rm soft}   (x )   %
 \ , 
 \
 \label{deltaQR}
\end{align} 
with  the kernel $R  (\eta;0) \equiv R(\eta)$  specified  by 
  \begin{align}
R  (\eta)  =  &  %
\int_{0}^1    \frac{du }{   
  |\eta -u  |}    B(u) 
 \  . 
 \
 \label{RfQ}
\end{align} 

At first sight,  one would expect a $\sim 1/ |\eta|$ behavior for large $|\eta|$
from Eq. (\ref{RfQ}). 
However, the $1/ |\eta|$  term is accompanied by the integral of $B(u)$ which vanishes  because 
of the plus-prescription structure of $B(u)$.  This is also the reason why   $\gamma_0$  in Eq. (\ref{gamman})  vanishes. 
Hence, in the  region $\eta >1$, we can write the  kernel as a series in $1/\eta$ starting with $n=1$,  
 \begin{align}
R(\eta) &|_{\eta >1}   = - \sum_{n=1}^\infty \frac{\gamma_{n}}{\eta^{n+1}} \ ,
\label{Routp}
\end{align}
or, in 
 a closed form  \cite{Radyushkin:2017lvu},
  \begin{align}
R(\eta) &|_{\eta >1} \equiv R_> (\eta) = \frac{1+\eta^2}{\eta-1}\ln
   \left(\frac{\eta-1}{\eta}\right) +
   \frac{3}{2 (\eta-1)}+1 \   . 
   \label{right}
\end{align}

Similarly, for  negative values, we have  the expansion 
 \begin{align}
R(\eta) &|_{\eta <-1}= \sum_{n=1}^\infty \frac{\gamma_{n}}{\eta^{n+1}} \ ,
\label{Routn}
\end{align}
and a closed-form expression  \cite{Radyushkin:2017lvu}
 \begin{align}
R(\eta) |_{\eta<0}\equiv R_< (\eta)   =&\frac{1+\eta^2 }{1-\eta}
\ln 
   \left(\frac{1-\eta}{-\eta}\right) +
   \frac{3}{2 (1-\eta)}-1 \   . 
   \label{left} 
 \end{align}
 
 \subsection{Large-$|y|$  behavior in QCD} 
 
According to Eq. (\ref{gamman}), we have 
 $\gamma_1 = 4/3$. Thus,   the asymptotic behavior for large $|\eta|$  is given by 
\begin{align}
R(\eta;0) |_{|\eta | \gg 1} =   -\frac43  \frac {{ \rm sgn} (\eta) }{\eta^2} +{\cal O} (1/\eta^3)   \  .
\end{align}
    
 The $\sim{{\rm sgn} (\eta) }/{\eta^2} $ behavior of $R(\eta)$ translates 
 into the \mbox{$\sim  {{ \rm sgn} (y) }/{y^2} $}   behavior of the quasi-PDF $Q(y,P)$ 
 for large values  of $|y|$.  As a result,  the $y^0$  moment  of $Q(y,P)$  converges
 for large $|y|$,  while  further   moments involve    divergences,
 in agreement  with observations made in \mbox{Sect. \ref{lowmom} }.
 In particular, the  $y^2$ moment involves a   linear divergence.
 If $B(u)$ would not  have the plus-prescription property,
 the divergence would be quadratic. 
  This agrees  with the estimate  made in 
 Sect. \ref{highmom}. 
 
 Hence, the divergences of the $y^n$ integrals   correspond to the presence 
of the  \mbox{$P$-independent}   terms   $\sim 1/y^2$   in the hard part of the quasi-PDFs $Q(y,P)$
 outside of the $0\leq y \leq 1$ region.

As we discussed, the $\ln z_3^2$  part of the  pseudo-ITDs contributes slowly-decreasing 
($\sim 1/y$ or $\sim 1/y^2$) terms into the 
$|y|>1$  part of quasi-PDFs. It is these terms that  lead to the divergence 
of the $y^n$  moments of the quasi-PDFs $Q(y,P)$.

 \subsection{Large-$P$ matching}
 
 These terms are not eliminated by  just taking 
 the $P \to \infty$ limit. 
 However, 
 they disappear 
 when one extracts $f(y,\mu^2)$  using  the matching condition (\ref{fNL}). Namely, we have 
   \begin{align} 
    f   (y, \mu^2)   = &    \widetilde Q(y,P)  +  \frac{\alpha_s}{2\pi} \, C_F\,  
\int_0^1  \frac{du}{u}  \,   f (y/u,\mu^2)   \nn &
 \times  \left  \{   B(u) \,  \ln \left (\mu^2/P^2 \right )  + C(u)  
   \right  \}   
+{\cal O} (1/P^2)
\   . 
\label{fNL2}
 \end{align}

 Since  both the ${\cal O} (1/P^2)$  soft part 
and the    $\widetilde Q(y,P)$  combination  of Eq. (\ref{QNL})  vanish for $|y|>1$  
  in the  \mbox{$P\to \infty$}  limit, Eq. (\ref{fNL2})  
 resolves  the problem of the support   mismatch    between $f(y,\mu^2)$ and  $Q(y,P)$. 
 As a result,  one  can calculate the $y^n$ moments 
 of the light-cone PDFs $f(y,\mu^2)$ using Eq. (\ref{fNL2})
 without getting divergences in its  right-hand side. 
 
As already noted, if we separate  quasi-PDFs  corresponding to 
 the real [$Q_- (y,P)$]  and imaginary  [$Q_+ (y,P)$]  parts of  the ITD, it is sufficient to consider positive $y$ only. 
Using the fact that perturbative part of $\widetilde Q(y,P)$ vanishes   outside the $|y|\leq 1$ region, 
we may write the   iterative solution of Eq. (\ref{fNL2}) for $y>0$ as  
   \begin{align} 
    f_\mp  &  (y, \mu^2)   =   Q_\mp (y,P) \, \theta (0\leq  y \leq 1)   \nn &   - \frac{\alpha_s}{2\pi} \, C_F\,  
  \int_0^1  \frac{dx}{x}  \, \,   
   [ Q_\mp  (x,P)  -Q_\mp  (y,P)  ]
 \nn & \times \Biggl  [  \theta (x \geq y)\left \{
 \frac{1+y^2/x^2}{1-y/x} \left ( \ln \left [4 y  (x-y) \frac{P^2}{\mu^2} \right ]  -1 \right ) +\frac{3/2}{ 1-y/x}  + 1
   \right\} 
 \nn &   + \theta (x \leq y) R_> (y/x)  \pm R_< (-y/x)
\Biggr  ]
+{\cal O} (1/P^2)
\   . 
\label{fNL3}
 \end{align}
Here the function $f_-(y)$   corresponds  to the real  part of the ITD
and  is given by $q(y)-\bar q (y)$, while $f_+(y)$  corresponds  to the  imaginary part of the ITD
 and  is given by $q(y)+\bar q (y)$. 
The kernels $R_> (\eta), R_< (\eta)$ are given by Eqs. (\ref{right}) and (\ref{left}).
The third line of Eq. (\ref{fNL3}) comes from $R^{\rm middle} (\eta)$ of Eq. (\ref{central})
and terms from Eq. (\ref{MNL0}). 
All the  terms explicitly written in Eq.  (\ref{fNL3})  involve 
quasi-PDFs in the $y<1$ region only. The $y>1$ part of  $\widetilde Q (y,P)$ is included in
${\cal O} (1/P^2)$ term and vanishes in the $P \to \infty$ limit.

We remind that  the   starting point for  the derivation of 
Eq. (\ref{fNL3}) is based on   Eqs.  (\ref{MNL0}) and (\ref{fNL}).
Hence, Eq. (\ref{fNL3})  applies to quasi-PDFs built from the {\it reduced}  pseudo-ITDs  (\ref{redm0}).

\section{Summary and conclusions}

In this paper, we discussed a specific feature of the quasi-PDFs
$Q(y,P_3)$ in which they differ from the usual PDFs $f(x)$,
namely, the presence of terms outside the $|y|\leq 1$ region. 

 In  a  model with a transverse momentum cut-off, such terms disappear in the $P \to \infty$ limit.
 However, in renormalizable theories, including QCD,
 one has  $|y|>1$ terms persisting (for a fixed  $\alpha_s$) even in the $P\to \infty$ limit.
 These terms have a  perturbative origin that may be traced to the $\ln z_3^2$ 
 singularities of the generating matrix element 
 $\langle p| \bar \psi (0)  \ldots  \psi (z_3) |p \rangle$.
 
 Since one {\it knows} that such terms, absent in  the light-cone PDFs  $f(x)$,  {\it must} 
 be present in the  quasi-PDFs $Q(y,P_3)$, one should just subtract them from 
  $Q(y,P_3)$ obtained on the lattice. The resulting ``reduced'' quasi-PDF 
  $\widetilde Q  (y,P)$
   for large $P$
  has support in the canonical  region  $|y|\leq 1$ only. 
  On a formal level, such a subtraction is automatically provided by 
  implementing the matching conditions.

Eq. (\ref{fNL3}),  that  is  given at the end of the paper,  
provides an explicit expression for  the lightcone PDF $f (y, \mu^2)$ involving the 
 quasi-PDF $Q(y,P)$ in the $|y| \leq 1$ region. Hence, 
in actual lattice PDF extractions, 
one may   ignore the $|y|>1$ region altogether and operate with $Q(y,P)$ obtained in the 
$|y| \leq 1$ region only.

A  related practical question is if the complications with
the  \mbox{$|y|>1$}  region  may be avoided?  
Indeed, according to the OPE (\ref{MNL0}), the reduced pseudo-ITD $  {\mathfrak  M} ( \nu, z_3^2) $,
a function directly ``coming  out of   the computer box'',   may be used, without intermediaries,
 to extract the
lightcone ITDs ${\cal I} (\nu, \mu^2)$. The latter are  the  Fourier transforms of the lightcone PDFs $f(x,\mu^2)$,
the functions that have the canonical  $|x|\leq 1$ support. 
Such an approach has been already applied in the exploratory 
lattice calculation  \cite{Orginos:2017kos} and in the 
construction \cite{Radyushkin:2018cvn} of $\overline{\rm MS}$ 
ITD  ${\cal I} (\nu, \mu^2)$ based on its results.

 \vspace{-3mm} 

\section*{Acknowledgements}

 \vspace{-1mm} 

I thank J.W.  Qiu for his interest in this work and discussions,
and W. Broniowski for correspondence. 
This work is supported by Jefferson Science Associates,
 LLC under  U.S. DOE Contract \#DE-AC05-06OR23177
 and by U.S. DOE Grant \#DE-FG02-97ER41028.



\begin{thebibliography}{10}


\bibitem{Feynman:1973xc}
  R.~P.~Feynman,
  ``Photon-hadron interactions,''
 W.A. Benjamin, Reading MA (1972), 282pp.



\bibitem{Ji:2013dva}
  X.~Ji,
  Phys.\ Rev.\ Lett.\  {\bf 110} (2013) 262002



\bibitem{Xiong:2013bka}
  X.~Xiong, X.~Ji, J.~H.~Zhang and Y.~Zhao,
  Phys.\ Rev.\ D {\bf 90} (2014) no.1,  014051



\bibitem{Lin:2014zya}
  H.~W.~Lin, J.~W.~Chen, S.~D.~Cohen and X.~Ji,
  Phys.\ Rev.\ D {\bf 91} (2015) 054510



\bibitem{Alexandrou:2015rja}
  C.~Alexandrou, K.~Cichy, V.~Drach, E.~Garcia-Ramos, K.~Hadjiyiannakou, K.~Jansen, F.~Steffens and C.~Wiese,
  Phys.\ Rev.\ D {\bf 92} (2015) 014502


\bibitem{Ma:2017pxb}
  Y.~Q.~Ma and J.~W.~Qiu,
  Phys.\ Rev.\ Lett.\  {\bf 120} (2018) no.2,  022003
  
\bibitem{Rossi:2017muf}
  G.~C.~Rossi and M.~Testa,
  Phys.\ Rev.\ D {\bf 96} (2017) no.1,  014507




\bibitem{Rossi:2018zkn}
  G.~Rossi and M.~Testa,
  Phys.\ Rev.\ D {\bf 98} (2018) no.5,  054028


\bibitem{Radyushkin:2017cyf}
  A.~V.~Radyushkin,
  Phys.\ Rev.\ D {\bf 96} (2017) no.3,  034025



\bibitem{Ioffe:1969kf}
  B.~L.~Ioffe,
  Phys.\ Lett.\  {\bf 30B} (1969) 123.



\bibitem{Braun:1994jq}
  V.~Braun, P.~Gornicki and L.~Mankiewicz,
  Phys.\ Rev.\ D {\bf 51} (1995) 6036



\bibitem{Radyushkin:2016hsy}
  A.~Radyushkin,
  Phys.\ Lett.\ B {\bf 767} (2017) 314





\bibitem{Radyushkin:1983wh}
  A.~V.~Radyushkin,
  Phys.\ Lett.\  {\bf 131B} (1983) 179.

\bibitem{Radyushkin:2017ffo}
  A.~Radyushkin,
  Phys.\ Lett.\ B {\bf 770} (2017) 514

\bibitem{Altarelli:1977zs}
  G.~Altarelli and G.~Parisi,
  Nucl.\ Phys.\ B {\bf 126} (1977) 298.
  
\bibitem{Gribov:1972ri}
  V.~N.~Gribov and L.~N.~Lipatov,
  Sov.\ J.\ Nucl.\ Phys.\  {\bf 15} (1972) 438
   [Yad.\ Fiz.\  {\bf 15} (1972) 781].
  

  
  
\bibitem{Dokshitzer:1977sg}
  Y.~L.~Dokshitzer,
  Sov.\ Phys.\ JETP {\bf 46} (1977) 641
   [Zh.\ Eksp.\ Teor.\ Fiz.\  {\bf 73} (1977) 1216].
  
\bibitem{Broniowski:2017gfp}
  W.~Broniowski and E.~Ruiz Arriola,
  Phys.\ Rev.\ D {\bf 97} (2018) no.3,  034031

\bibitem{Alexandrou:2016eyt}
  C.~Alexandrou, K.~Cichy, M.~Constantinou, K.~Hadjiyiannakou, K.~Jansen, F.~Steffens and C.~Wiese,
  PoS LATTICE {\bf 2016} (2016) 151

\bibitem{Ji:2017rah}
  X.~Ji, J.~H.~Zhang and Y.~Zhao,
  Nucl.\ Phys.\ B {\bf 924} (2017) 366



\bibitem{Polyakov:1980ca}
  A.~M.~Polyakov,
  Nucl.\ Phys.\ B {\bf 164} (1980) 171.



\bibitem{Ishikawa:2017faj}
  T.~Ishikawa, Y.~Q.~Ma, J.~W.~Qiu and S.~Yoshida,
  Phys.\ Rev.\ D {\bf 96} (2017) no.9,  094019



\bibitem{Ji:2017oey}
  X.~Ji, J.~H.~Zhang and Y.~Zhao,
  Phys.\ Rev.\ Lett.\  {\bf 120} (2018) no.11,  112001



\bibitem{Green:2017xeu}
  J.~Green, K.~Jansen and F.~Steffens,
  Phys.\ Rev.\ Lett.\  {\bf 121} (2018) no.2,  022004
  



\bibitem{Izubuchi:2018srq}
  T.~Izubuchi, X.~Ji, L.~Jin, I.~W.~Stewart and Y.~Zhao,
  Phys.\ Rev.\ D {\bf 98} (2018) no.5,  056004
  

\bibitem{Radyushkin:2017lvu}
  A.~V.~Radyushkin,
  Phys.\ Lett.\ B {\bf 781} (2018) 433
  
  
\bibitem{Radyushkin:2018cvn}
  A.~Radyushkin,
  Phys. \ Rev.  \ D  {\bf 98}  (2018)  014019. 




\bibitem{Orginos:2017kos}
  K.~Orginos, A.~Radyushkin, J.~Karpie and S.~Zafeiropoulos,
  Phys.\ Rev.\ D {\bf 96} (2017) no.9,  094503.
  



  

  
  
  \end{thebibliography}
\end{document}